\documentclass[journal=jcp,manuscript=article]{achemso}
\setkeys{acs}{maxauthors=0}

\usepackage{achemso}
\usepackage{graphics}
\usepackage{amssymb,amsfonts}
\usepackage{graphicx}
\usepackage[table]{xcolor}
\usepackage{caption}
\usepackage{subcaption}
\usepackage{colortbl}
\usepackage{amsmath}
\usepackage{amsopn}
\usepackage{bm}
\usepackage{color}
\usepackage{array}
\usepackage{lscape}
\usepackage{mciteplus}
\usepackage[version=3]{mhchem}
\usepackage{ulem}
\usepackage{listings}
\usepackage{fancyvrb}
\usepackage{enumerate}
\SectionNumbersOn
\makeatletter
\newcommand*\rel@kern[1]{\kern#1\dimexpr\macc@kerna}
\newcommand*\widebar[1]{%
  \begingroup
  \def\mathaccent##1##2{%
    \rel@kern{0.8}%
    \overline{\rel@kern{-0.8}\macc@nucleus\rel@kern{0.2}}%
    \rel@kern{-0.2}%
  }%
  \macc@depth\@ne
  \let\math@bgroup\@empty \let\math@egroup\macc@set@skewchar
  \mathsurround\z@ \frozen@everymath{\mathgroup\macc@group\relax}%
  \macc@set@skewchar\relax
  \let\mathaccentV\macc@nested@a
  \macc@nested@a\relax111{#1}%
  \endgroup
}
\makeatother
\numberwithin{equation}{subsection}

\fvset{frame=single,framesep=1mm,fontfamily=courier,fontsize=\footnotesize,framerule=.3mm,commandchars=\\\{\}}

\author{Janus J. Eriksen}
\email{jeriksen@uni-mainz.de}
\affiliation[Johannes Gutenberg-Universit\"at Mainz]{Institut f\"ur Physikalische Chemie, Johannes Gutenberg-Universit\"at Mainz, D-55128 Mainz, Germany}

\title[TITLE]
  {Efficient and portable acceleration of quantum chemical many-body methods in mixed floating point precision using OpenACC compiler directives}

\begin{document}
%
%
\begin{abstract}
%

%
It is demonstrated how the non-proprietary OpenACC standard of compiler directives may be used to compactly and efficiently accelerate the rate-determining steps of two of the most routinely applied many-body methods of electronic structure theory, namely the second-order M\o ller-Plesset (MP2) model in its resolution-of-the-identity (RI) approximated form and the (T) triples correction to the coupled cluster singles and doubles model (CCSD(T)). By means of compute directives as well as the use of optimized device math libraries, the operations involved in the energy kernels have been ported to graphics processing unit (GPU) accelerators, and the associated data transfers correspondingly optimized to such a degree that the final implementations (using either double and/or single precision arithmetics) are capable of scaling to as large systems as allowed for by the capacity of the host central processing unit (CPU) main memory. The performance of the hybrid CPU/GPU implementations is assessed through calculations on test systems of alanine amino acid chains using one-electron basis sets of increasing size (ranging from double- to pentuple-$\zeta$ quality). For all but the smallest problem sizes of the present study, the optimized accelerated codes (using a single multi-core CPU host node in conjunction with six GPUs) are found to be capable of reducing the total time-to-solution by at least an order of magnitude over optimized, OpenMP-threaded CPU-only reference implementations.

\end{abstract}
\newpage
%

%
%
\section{Introduction}\label{intro_section}

When formulating the criteria for what defines {\it{portable}} source code, one tends to draw a distinction between two types of portability, namely that associated with functionality and that associated with performance. While the former of these covers what is perhaps usually implied when referring to code portability, i.e., the ability of a single code to run anywhere (rebuild the code on various architectures, the code will run and produce correct results), the latter, the existence of which have often been declared a fallacy~\bibnote{See, for instance, the talk by Jeff Larkin on performance portability from the SC15 conference: {\url{http://images.nvidia.com/events/sc15/pdfs/SC15-Performance-Portability-OpenACC-Larkin.pdf}} (accessed \today)}, relates to the ability of a single code to run productively anywhere, i.e., achieving, say, $80-90 \%$ of the hand-tuned performance on any given architecture. As such, making a code portable will often result in a compromise between superior performance on a single platform and decent performance on all potential platforms. One possible way of circumventing this problem is by cluttering the source code with multiple conditional preprocessor directives and retain architecture-specific versions of key kernels with vendor-specific intrinsics. However, pursuing this strategy is bound to severely hamper two of the other main best practices of code development, namely those which are concerned with securing code {\it{maintainability}} and {\it{testability}}~\cite{best_practices_plos_biol_2014}.\\

For execution on standard multi-core central processing unit (CPU) architectures, the {\it{OpenMP standard}}~\cite{openmp_website}---first released in the second half of the 1990s as a means to parallelize loops across compute threads---has nowadays manifested itself as the de facto programming model for intranode worksharing and, in particular, single-instruction multiple-data (SIMD) parallel instructions. The advantages of employing OpenMP compiler directives are manifold; the standard is open and supported by most vendors, the compiler directives (or pragmas) are ignored entirely by those few compilers that do not offer support, and the syntax is relatively simple to comprehend, even to the uninitiated, making the learning barrier less steep than, e.g., that for intra- and internode message passing using the MPI protocol. Recently, with the advent of OpenMP versions 4.0 and 4.5, the standard now also allows for targeting other types of architectures such as many-core x86 processors, e.g., Intel\textsuperscript{\textregistered} Xeon Phi\textsuperscript{\texttrademark} products, and graphics processing units (GPUs), e.g., NVIDIA\textsuperscript{\textregistered} or AMD\textsuperscript{\textregistered} GPUs. These feature enhancements from homo- to heterogeneous platforms allow for offloading of compute intensive code regions to coprocessors via so-called {\it{device}} constructs (in this terminology, the CPU, to which the coprocessor is connected, is denoted the {\it{host}}). However, there exist intrinsic difficulties in transferring the worksharing capabilities of OpenMP on the host to a corresponding worksharing on any attached device, in particular GPUs, as the architectural details of these are fundamentally different from those of the host. Specifically, the traditional OpenMP model of spawning a single team of threads, which then divide loop iterations among them, will not be suitable per se for execution on GPUs, as these are composed of multiple processing elements (PEs) that run in parallel (with no synchronization options between the threads in different PEs) and where each individual PE has the ability to efficiently perform vector-like operations. There are ways to bypass this issue, though, for instance by creating multiple teams of threads, across which the workload may be distributed. However, this is almost guaranteed to result in {\it{accelerated}} code that will not execute satisfactorily on the host (thereby losing performance portability), which will reintroduce the need for {\bf{(i)}} preprocessing of different code blocks in addition to {\bf{(ii)}} code duplication with the required maintenance that inevitably arises as a result of this~\bibnote{See, for instance, the talk by James Beyer and Jeff Larkin on OpenMP vs. OpenACC from the GTC2016 conference:\\{\url{http://on-demand.gputechconf.com/gtc/2016/presentation/s6410-jeff-larkin-beyer-comparing-open-acc-openmp.pdf}} (accessed \today)}.\\

Instead, an alternative standard---the {\it{OpenACC standard}}~\cite{openacc_website}, recently founded by a consortium consisting of NVIDIA\textsuperscript{\textregistered} and several key vendors, including Cray\textsuperscript{\textregistered} and PGI\textsuperscript{\textregistered}---has been established in an attempt to facilitate such portable and maintainable parallelism across various architectures, that is, provide the ability to expose to the compiler in a simple way what parts of the code can be run in parallel, and then leave it to the compiler to build the accelerated code {\it{on a given platform}}, be that the host (as a real alternative to OpenMP) or the device (as an inherently different approach to OpenMP). Since OpenACC is designed with heterogeneous programming in mind, it is not biased towards multi-core CPU architectures in the same way as OpenMP is, while still being able to advantage from past endeavours made by compiler developers that tuned these towards vectorization. In particular, as opposed to adhering to a strictly prescriptive programming model of simple thread teams, the OpenACC standard extends this by adding an additional layer of flexibility; for instance, when targeting GPUs, on which it is reasonable to think of a PE as a streaming multiprocessor (SM), the standard offers the possibility of parallelizing over threadblocks in a grid, i.e., individual SMs, warps (a group of 32 threads comprising the minimum execution unit), as well as individual CUDA threads within a warp. In the OpenACC terminology, these three layers are known as gangs, workers, and vectors.\\

The OpenACC programming model is a so-called host-directed sequential model capable of leveraging the parallel potential of one or even multiple accelerator devices, be that GPUs or multi-core coprocessors. The constructs, which the standard is designed to target, are the same as those the OpenMP standard are aimed at, namely nested loop structures, which themselves are suitable for shared memory multiprocessing. The translator between the input program and the individual SIMD units on the target architecture is the compiler, and the vectorization inhibitors are thus the same as for OpenMP, e.g., loop carried dependencies, prohibitive complexity (e.g., function calls), as well as indirect addressing and striding. As the compiler is directed by pragmas used to expose the parallel regions in the code, the code developer may approach the problem of offloading a certain kernel in a descriptive and, importantly, incremental manner, which makes for a less cumbersome workflow than if the entire code has to be implemented from scratch using some low-level approach unrelated to the language in which the actual code base is written (e.g., CUDA or OpenCL, in a GPU context). Furthermore, as the accelerated code will be based on the original source, this warrants, in turn, a functionally and performance portable, intuitively transparent, and easily extendable and maintainable implementation.\\

In the present work, it will be demonstrated how to make use of OpenACC compiler directives in the acceleration of electronic structure theories, with an illustrative application to two popular many-body methods, namely the second-order M\o ller-Plesset (MP2) model~\cite{mp2_phys_rev_1934} in its resolution-of-the-identity (RI) approximated form and the preeminent perturbative (T) triples correction to the coupled cluster~\cite{cizek_1,*cizek_2,*paldus_cizek_shavitt} (CC) singles and doubles (CCSD) model~\cite{ccsd_paper_1_jcp_1982}, compositely known as the CCSD(T) model~\cite{original_ccsdpt_paper,ccsdpt_perturbation_stanton_cpl_1997}. To focus the discussion, the present work will be solely concerned with offloading to GPU accelerators, although we stress once again how the OpenACC standard, and hence the present approach, is completely general and, as such, not limited to nor biased against accelerators of a specific kind. On that note, we will start by briefly alluding to the existing literature, which---despite the use of general-purpose GPUs in natural sciences being a rather recent topic---is rich with work devoted to GPU-accelerated quantum chemistry, concerned with diverse topics ranging from the generation of electron repulsion integrals (ERIs)~\cite{ufimtsev_martinez_gpu_eri_jctc_2008,*luehr_martinez_gpu_eri_jctc_2011,asadchev_gordon_gpu_eri_jctc_2010,gamess_uk_gpu_eri_jcc_2011,miao_merz_gpu_eri_jctc_2013,*miao_merz_gpu_eri_jctc_2015,toivanen_sundholm_gpu_gbfmm_pccp_2015}, over self-consistent Hartree-Fock (HF), complete active space (CAS), and density functional theory (DFT) methods~\cite{yasuda_gpu_dft_jctc_2008,ufimtsev_martinez_gpu_scf_jctc_2009,*ufimtsev_martinez_gpu_gradient_jctc_2009,asadchev_gordon_gpu_hf_jctc_2012,hohenstein_martinez_gpu_casscf_jcp_2015_1,*hohenstein_martinez_gpu_casscf_jcp_2015_2,*snyder_martinez_gpu_casscf_jpcl_2016,yoshikawa_nakai_gpu_dac_scf_jcc_2015,andrade_aspuru_guzik_gpu_dft_jctc_2013,cawkwell_mniszewski_dens_matrix_jctc_2014,losilla_sundholm_gpu_fock_jctc_2015,kussmann_ochsenfeld_gpu_exchange_jctc_2015}, to solvent models~\cite{liu_martinez_gpu_imd_jctc_2015,nitsche_lebrero_gpu_dft_qmmm_jctc_2014}, force fields~\cite{amber_gpu_1_jctc_2012,*amber_gpu_2_jctc_2013}, and semi-empirical~\cite{wu_thiel_gpu_semiempirical_jctc_2012} as well as many-body methods~\cite{vogt_aspuru_guzik_gpu_rimp2_jpca_2008,*olivares_aspuru_guzik_gpu_mgemm_rimp2_jctc_2010,*watson_aspuru_guzik_gpu_mgemm_rimp2_cse_2010,nakajima_gpu_rimp2_jcc_2016,hirata_gpu_mp2_jctc_2016,deprince_hammond_gpu_ccsd_jctc_2011,deprince_sherrill_gpu_ccsd_mol_phys_2014,asadchev_gordon_gpu_ccsdpt_jctc_2013,leang_gordon_gpu_matrix_jctc_2014,ma_kowalski_gpu_tensor_clust_comput_2013,ma_kowalski_gpu_reg_ccsdpt_jctc_2011,*bhaskaran_nair_kowalski_gpu_mr_cc_jctc_2013,fales_levine_gpu_fci_jctc_2015}. Adding to this quantum, several program packages are today released with full or partial GPU-enabled features, e.g., Terachem~\cite{terachem}, NWChem~\cite{nwchem}, among others.\\

That said, the adaption of GPUs and other types of coprocessors in the acceleration of quantum chemical and physical methods and applications as a whole has often been criticized for being too difficult a task to undertake implementation-wise and hence not worth the effort, in part also because of the uncertainty of (the shape of/general consensus on) future architectures that have always permeated the community (hesitation towards vectorization, massive parallelism, high-performance computing, etc.)~\cite{pratx_xing_med_phys_2011}. On top of that, it is important to note how the clear majority of all quantum chemical implementations in existence today have been developed and are being maintained by domain scientists, many of which hold temporary positions working (simultaneously) on various theoretical and/or application-oriented projects~\cite{jacob_open_source_software_jpcl_2016}. Thus, while said scientists might decide to implement a given accelerated method from scratch (again, presumably making use of some low-level approach), they might not be responsible for extending it with new features and capabilities in the future nor for the routine maintenance required with the emergence of novel architectures, not to mention the finer architectural diversities between different generations of, for instance, GPU accelerators. Add to that the typical requirements of a single code base and, in particular, platform independence, which most codes are subject to in the sense that any addition of accelerated code must not interfere with the standard compilation process (cf. the discussion in the opening paragraph of the present Section), and one will potentially arrive at more reasons against investing the necessary efforts than actual reasons in favor of doing so.\\

However, a recent paper from the Mart{\'i}nez group at Stanford University has outlined an alternative strategy for avoiding the above issues, albeit one targeted exclusively at GPU hardware~\cite{song_martinez_gpu_ace_jctc_2016}. The paper, which succeeds earlier work on so-called meta-programming from within the same group~\cite{titov_martinez_gpu_meta_prog_jctc_2013}, proposes the application of an automatic code generator capable of employing mathematical knowledge of the underlying algorithms (for the efficient generation of ERIs, in this case) to create a wide spectrum of possible code variants that explore, in their own words, the full space of all possible program implementations. This type of empirical, so-called profile-guided optimization subsequently tests the (multiple) candidate implementations on the target architecture at hand before deciding upon a final version, and the combination of a graph-based code generator, a tester, and an optimizer hence aims at producing the optimal implementation for previous and current generations of GPUs as well as, at least in principle, future generations.\\

These numerous code variants, which may be automatically produced using the engine in Ref. \citenum{song_martinez_gpu_ace_jctc_2016}, will each make up different compromises between a number of parameters and factors; for instance, the efficiency of a given kernel will be tightly bound to its utilization of the complex memory hierarchy on the GPUs, the latency of which increases drastically upon traversing up through it, and the algorithmic complexity of the individual kernels trivially increases with respect to the maximum angular momentum of the one-electron basis set. Furthermore, the choice of resulting implementation will also depend on the choice of floating point precision, since lowering this from double to single precision---granted that the requested accuracy allows for it---may give preference to other variants. Importantly, however, the use of single precision arithmetics will result in an overall speed-up of the calculation, for execution on the host CPU and particularly so for execution on GPUs. Now, in comparison with the generation of ERIs, the computation of correlation energies using (perturbative) many-body methods poses rather different requirements on the final implementation, as we will return to in \ref{implementations_section}. That being said, it is the intent of the present author to illustrate, through OpenACC implementations of the RI-MP2 and CCSD(T) methods, how simple and transparent such methods may be accelerated. It should be noted, however, that the adaption of an existing CPU ERI code to GPUs (or other types of accelerators, {\it{vide supra}}) by means of OpenACC directives will no doubt be more challenging, but it is still probably easier for the developer than using, e.g., CUDA and writing the code from scratch. In particular, the programmer may start by incrementally porting one kernel after another by exposing the inherent parallelism (loop structures) of the algorithm using directives, thereby verifying the correctness of the implementation along the way prior to subjecting the initial implementation to a subsequent optimization. In addition, the final implementation will require a minimum of maintenance, as the optimizer---with respect to novel architectures and compute capabilities---remains the compiler itself, and since the directives are treated as mere comments to a non-accelerating compiler, this applies for the efforts that will have to be invested into platform independence of the source as well.\\

The present work is organized as follows. Following a brief introduction to the RI-MP2 and CCSD(T) methods in \ref{theory_section}, the actual OpenACC implementations of these---using a combination of single and double precision arithmetics---are discussed in \ref{implementations_section}. In \ref{com_details_section}, we provide some computational details on the calculations to follow in \ref{results_section}, in which the performance of the implementations is assessed through calculations on test systems consisting of alanine amino acids in $\alpha$-helix arrangements using one-electron basis sets of increasing size (ranging from double- to pentuple-$\zeta$ quality). Finally, a short summary alongside some conclusive remarks are presented in \ref{conclusion_section}.

%
%
\section{Theory}\label{theory_section}

In electronic structure theory, CC and many-body perturbation theory~\cite{shavitt_bartlett_cc_book} (MBPT) both offer a systematic approach towards the full configuration interaction~\cite{mest} (FCI) wave function---the exact solution to the time-independent, non-relativistic electronic Schr\"odinger equation within a given one-electron basis set. In both hierarchies of methods, the mean-field HF solution acts as the reference to which correlated corrections are made by including excited configurations in the wave function. Moving beyond the HF approximation, not only Fermi correlation is considered, as also the correlated motion of electrons of opposite spin start being described. This implies that upon traversing up through either of the CC or MBPT hierarchies, increasingly more of such dynamical correlation is included in the wave function, with a complete description met at the target FCI limit.\\

In the MBPT series, the lowest-order perturbative correction to the HF energy for the isolated effect of connected double excitations is that of the non-iterative MP2 model~\cite{mp2_phys_rev_1934}. As an improvement, the CC hierarchy offers the iterative CCSD model~\cite{ccsd_paper_1_jcp_1982}, which accounts not only for connected double excitations, but does so to infinite order in the space of all single and double excitations out of the HF reference. However, in order to achieve relative energies of chemical accuracy ($\sim 1$ kcal/mol), the CCSD energy has to be augmented by correction terms built from higher-level excitations. To a lowest approximation, corrections for effects due to connected triple excitations have to be considered. Among such higher-level approaches, the CCSD(T) model~\cite{original_ccsdpt_paper} is by far the most prominent. Unfortunately, the additional accuracy of the CCSD and, in particular, the CCSD(T) models over the MP2 model comes at a price, with the evaluation of the rate-determining step scaling non-iteratively as $\mathcal{O}(N^5)$ in the MP2 model (where $N$ is a composite measure of the total system size), iteratively as $\mathcal{O}(N^6)$ in the CCSD model, and non-iteratively as $\mathcal{O}(N^7)$ for the (T) correction to the converged CCSD energy. Thus, whereas all three methods have nowadays become standard tools to computational quantum chemists, the current application range of the MP2 model considerably exceeds that of the CCSD(T) model, which are the two models that we will limit our focus to herein, due to the practical differences in associated computational cost.

%
%
\subsection{RI-MP2}\label{rimp2_theory_subsection}

In spite of the fact that the MP2 model is formally far less expensive than the CCSD(T) model, the time-to-solution for the average MP2 calculation may still be substantial when targeting increasingly larger systems, e.g., in comparison with cheaper, albeit less rigorous and less systematic methods such as those of semi-empirical nature or those that calculate the energy from a functional of the one-electron density (DFT methods). For this reason, most efficient implementations of the MP2 model invoke an RI approximation for reducing the computational cost (the prefactor) as well as lowering the memory constraints~\cite{whitten_jcp_1973,feyereisen_rimp2_cpl_1993,vahtras_ri_cpl_1993,weigend_haser_rimp2_tca_1997}. Despite being an approximation to the MP2 model, the focused development of optimized auxiliary basis sets has managed to significantly reduce the inherent RI error, and for most reasonable choices of the intrinsic thresholds of a modern implementation, the error affiliated with the actual approximation will be negligible. However, while the computational cost is notably reduced in an RI-MP2 calculation with respect to its MP2 counterpart, its formal $\mathcal{O}(N^5)$ scaling with the size of the system will often still deem it demanding if no fundamental algorithmic changes are made~\cite{almlof_cpl_1991,ayala_scuseria_rimp2_jcp_1999,werner_manby_knowles_rimp2_jcp_2003,doser_ochsenfeld_rimp2_jcp_2009,*doser_ochsenfeld_rimp2_zpc_2010,baudin_dec_rimp2_jcp_2016}.\\

In the closed-shell (canonical) MP2 model, the correlation energy is given by the following expression
\begin{align}
E_{\text{MP2}} = -\sum_{ab}\sum_{ij}\frac{g_{aibj}(2g_{aibj} - g_{ajbi})}{\epsilon^{ab}_{ij}} \label{mp2_energy}
\end{align}
in terms of two-electron ERIs, $g_{aibj}$ (Mulliken notation), over spatial (spin-free) HF virtual orbitals $a,b$ and occupied orbitals $i,j$, as well as the difference in energy between these
\begin{align}
\epsilon^{ab}_{ij} = \epsilon_a + \epsilon_b - (\epsilon_i + \epsilon_j) \ . \label{e_orb_diff}
\end{align}
Besides the final evaluation of the MP2 energy in \ref{mp2_energy}, the dominant part of an MP2 calculation is the construction of the four-index ERIs. In the RI-MP2 method, these are approximated by products of two-, $V_{\alpha\gamma}$, and three-index, $W_{ai\alpha}$, integrals by means of the following symmetric decomposition
\begin{align}
g_{aibj} \approx \sum_{\gamma}C^{\gamma}_{ai}C^{\gamma}_{bj} \ . \label{eri_sym_decomp}
\end{align}
In \ref{eri_sym_decomp}, Greek indices denote atomic orbitals within an auxiliary fitting basis set used for spanning the RI, and the fitting coefficients, $C^{\gamma}_{ai}$, are defined as
\begin{align}
C^{\gamma}_{ai} = \sum_{\alpha}W_{ai\alpha}[\textbf{V}^{-1/2}]_{\alpha\gamma} \ . \label{fitting_coef}
\end{align}
In terms of computational cost, the evaluation of two- and three-index integrals and/or the calculation of the fitting coefficients will dominate the overall calculation for small systems ($\mathcal{O}(N^3)$- and $\mathcal{O}(N^4)$-scaling processes, respectively). However, upon an increase in system size, the final $\mathcal{O}(N^5)$-scaling assembly of the two-electron integrals in \ref{eri_sym_decomp} will start to dominate, and this process is thus an ideal candidate for accelerating the RI-MP2 model, cf. \ref{implementations_section}. 

%
%
\subsection{CCSD(T)}\label{ccsdpt_theory_subsection}

As touched upon in the opening paragraph of the present Section, the (T) correction scales non-iterative as $\mathcal{O}(N^7)$, as opposed to the $\mathcal{O}(N^6)$-scaling iterative evaluation of the energy and cluster amplitudes in the preceding CCSD calculation. Thus, for all but the smallest systems, the evaluation of the (T) correction, at least in its conventional canonical formulation, will dominate. In a basis of local HF orbitals, however, this balance might shift, making the underlying evaluation of the CCSD energy (and amplitudes) the computational bottleneck~\cite{werner_schutz_cpl_2000,*schutz_jcp_2000,*werner_schutz_jcp_2011,friedrich_dolg_jctc_2009,kobayashi_nakai_jcp_2009,li_piecuch_jcp_2009, rolik_kallay_jcp_2011,*rolik_kallay_jcp_2013,schutz_manby_werner_jcp_2013,riplinger_neese_jcp_2013,eriksen_dec_ccsdpt_jctc_2015}.\\

For closed-shell systems, the CCSD(T) energy is defined by a triples correction, $E_{\text{(T)}}$, to the CCSD energy, in turn defined by two energy contributions rationalized from MBPT: $E^{[4]}$, a fourth-order term involving CCSD doubles amplitudes, $\{t^{ab}_{ij}\}$, and $E^{[5]}$, a fifth-order term involving CCSD singles amplitudes, $\{t^{a}_{i}\}$. In a basis of canonical HF spatial orbitals, these may be expressed as~\cite{piecuch_comp_phys_comm_2002}
\begin{subequations}
\label{piecuch_4th_and_5th_order}
\begin{align}
E^{[4]} &= \sum_{abc}\sum_{ijk}\tilde{t}^{abc}_{ijk}t^{abc}_{ijk}\epsilon^{abc}_{ijk} \label{piecuch_4th_order} \\
E^{[5]} &= \sum_{abc}\sum_{ijk}\tilde{z}^{abc}_{ijk}t^{abc}_{ijk}\epsilon^{abc}_{ijk} \ . \label{piecuch_5th_order} 
\end{align}
\end{subequations}
In \ref{piecuch_4th_and_5th_order}, $\epsilon^{abc}_{ijk}$ is defined on par with $\epsilon^{ab}_{ij}$ in \ref{e_orb_diff}, and the triples amplitudes, $\{t^{abc}_{ijk}\}$, are defined as
\begin{align}
t^{abc}_{ijk} = -P^{abc}_{ijk}\frac{\sum_{d}t^{ad}_{ij}g_{ckbd} - \sum_{l}t^{ab}_{il}g_{cklj}}{\epsilon^{abc}_{ijk}} \label{analytical_triples_ampl}
\end{align}
where $P^{abc}_{ijk}$ is a symmetrization operator
\begin{align}
P^{abc}_{ijk}x^{abc}_{ijk} = x^{abc}_{ijk} + x^{acb}_{ikj} + x^{bac}_{jik} + x^{bca}_{jki} + x^{cab}_{kij} + x^{cba}_{kji} \ . \label{symm_operator}
\end{align}
The $z^{abc}_{ijk}$ coefficients in \ref{piecuch_5th_order} are given as
\begin{align}
z^{abc}_{ijk} = -\frac{(t^{a}_{i}g_{jbkc} + t^{b}_{j}g_{iakc} + t^{c}_{k}g_{iajb})}{\epsilon^{abc}_{ijk}} \label{piecuch_z_coefficients}
\end{align}
and an arbitrary six-index quantity $\tilde{x}^{abc}_{ijk}$ in \ref{piecuch_4th_and_5th_order} is defined as
\begin{align}
\tilde{x}^{abc}_{ijk} = 2(\tfrac{2}{3}x^{abc}_{ijk} - x^{acb}_{ijk} + \tfrac{1}{3}x^{bca}_{ijk}) \ . \label{piecuch_tilde_quantity}
\end{align}
The rate-determining step is now identified as being the construction of the triples amplitudes in \ref{analytical_triples_ampl}, whereas the final evaluation of $E_{\text{(T)}}$ only scales as $\mathcal{O}(N^6)$. As opposed to the evaluation of the RI-MP2 energy in \ref{rimp2_theory_subsection}, for which only parts of the algorithm will be accelerated (\ref{mp2_energy} and \ref{eri_sym_decomp} combined), all of the (T) kernels will be offloaded to the accelerator(s), as will be detailed in the following \ref{implementations_section}.

%
%
\section{Implementations}\label{implementations_section}

As is obvious from \ref{theory_section}, the evaluation of both the RI-MP2 energy and the (T) correction relies on a vast number of matrix-matrix multiplications (MMMs), which in all modern implementations of the methods are handled by making calls to the {\texttt{dgemm}} routine of an optimized level-3 basic linear algebra subprograms~\cite{blas_original_paper_1,*blas_original_paper_2,*blas_original_paper_3} (BLAS-3) library. The fact that both methods may be recasted such that they utilize BLAS libraries optimized for modern multi-core CPU architectures (e.g., Intel\textsuperscript{\textregistered} MKL, ATLAS~\cite{atlas_blas_paper}, or GotoBLAS~\cite{gotoblas_paper}, to name a few) implies that they are amenable to accelerators as well (GPUs, in the present context), for which optimized BLAS libraries too exist, e.g., MAGMA~\cite{magma_blas_1,*magma_blas_2} and CUBLAS~\cite{cublas_website}. Thus, in contrast to the generation of ERIs, cf. the discussion in \ref{intro_section}, which is comprised of extensive parallel regions, each with numerous tightly nested loop structures, this forces somewhat different requirements on an accelerated implementation for coprocessors; in particular, aspects such as efficient data movement patterns and data reuse become highly important, since the tensors involved in many-body methods tend to grow large (in excess of what might be afforded by the available device memory) and the current interconnects between a host and its attached accelerator(s) (e.g., a PCIe bus between a CPU host and a GPU) generally impede unnecessary (or redundant) movement of data. The few remaining loop constructs ({\it{vide infra}}) are then translated by descriptive directives into kernels that can run in parallel on the accelerator. By letting the host processor orchestrate the execution of all parallel kernels, those cores on the host that do not control an accelerator are free to participate in the workload. This is an important feature; by allowing for the final code to execute in a hybrid manner on both the host CPU as well as the attached accelerator device(s), the present approach is not biased towards neither the former nor the latter of the two platforms (making for a fair comparison of the relative CPU vs. GPU throughput~\cite{lee_intel_debunking_paper_2010}). If the implementation was instead formulated in terms of some low-level language such as CUDA, such a bias would be present. In addition, the OpenACC standard further provides features for circumventing the default host-directed behavior, in which a sequence of operations gets scheduled for sequential execution on the device, by instead allowing for these to overlap in different pipelines, i.e., executing operations (computations and data transfers) asynchronously, given that all potential race conditions have been appropriately mapped out.\\

Throughout the years, elaborate attempts have been made at pushing the limits for the RI-MP2~\cite{bernholdt_harrison_par_rimp2_ijqc_2009,katouda_nagase_par_rimp2_ijqc_2009,katouda_nakajima_par_rimp2_jctc_2013,maschio_par_rimp2_jctc_2013} and CCSD(T)~\cite{molpro_ccsdpt_wires, aces_iii_ccsdpt_1,*aces_iii_wires, cfour_ccsdpt, pqs_ccsdpt_1,*pqs_ccsdpt_wires, gamess_ccsdpt_1,*gamess_ccsdpt_2, nwchem_ccsdpt_1,*nwchem_ccsdpt_2,nwchem_h2o_20_2} methods by devising massively parallel implementations of the standard canonical formulations in \ref{theory_section}. In addition, the methods have previously been ported to GPUs within the groups of Al{\'a}n Aspuru-Guzik~\cite{vogt_aspuru_guzik_gpu_rimp2_jpca_2008,*olivares_aspuru_guzik_gpu_mgemm_rimp2_jctc_2010,*watson_aspuru_guzik_gpu_mgemm_rimp2_cse_2010} (RI-MP2), Nakajima~\cite{nakajima_gpu_rimp2_jcc_2016} (RI-MP2), Mark Gordon~\cite{asadchev_gordon_gpu_ccsdpt_jctc_2013} (CCSD(T)), and Karol Kowalski~\cite{ma_kowalski_gpu_reg_ccsdpt_jctc_2011,*bhaskaran_nair_kowalski_gpu_mr_cc_jctc_2013} (CCSD(T)), as, e.g., reviewed in recent contributions to a book devoted to the subject of electronic structure calculations on graphics units~\cite{olivares_amaya_aspuru_guzik_gpu_rimp2_book_2016,ma_kowalski_gpu_ccsdpt_book_2016}. In most, if not all of these implementations, the main driver routine, regardless of the method, consists of an outer nested loops, most often over occupied orbital indices with restricted summations to incorporate permutational symmetry of the two-electron ERIs~\cite{rendell_aijkbc_algorithm_1,rendell_aijkbc_algorithm_2}. For both methods, each loop cycle then involves (numerous) large MMMs, possibly intertwined by tensor (index) permutations and the division by orbital energy differences. Since each loop cycle is data independent of all others, these may be trivially distributed over multiple CPU nodes using MPI.\\

In a likeminded fashion, OpenMP worksharing directives may be used to parallelize the individual loop cycles over the available ($N_{\text{threads}}$, or more generally, {\texttt{OMP$\_$NUM$\_$THREADS}}) cores on the host CPU node. In this case, $N_{\text{threads}}$ single-threaded MMMs are executed concurrently, which will generally be more efficient than running a single multi-threaded MMM (over $N_{\text{threads}}$ threads) sequentially $N_{\text{threads}}$ times, albeit at the expense of having to store multiple private copies of some of the involved intermediates. As a prerequisite for using OpenMP this way, one needs to flatten (or collapse) the two (in the case of the RI-MP2 method, cf. \ref{mp2_energy}) or three (in the case of the (T) correction, cf. \ref{piecuch_4th_and_5th_order}) outer loops into one large, composite outer loop with loop cycles of identical weight. This way, the evaluation of the RI-MP2 energy or (T) correction is programmed to execute in a hybrid heterogeneous manner on both the CPU (host) cores and one or even multiple accelerator devices; combined, the accelerators may be regarded as augmenting the host CPU by additional cores (one for each attached accelerator), albeit some that operate at a significantly higher clock rate. As far as the use of the GPUs is concerned, all of the required data blocks (tiles of the fitting coefficient tensor in the RI-MP2 code, and tiles of the involved CCSD doubles amplitudes and ERIs in the CCSD(T) code) can be transferred asynchronously, thereby overlapping all data movement between the host and the device(s) with computations on the latter, and for the actual tensors, we allocate these directly into page-locked (or pinned) memory in order to lower the cost of the data transfers.\\

Recently, in an invited chapter to a new book devoted to parallel programming with OpenACC, the present author outlined in some detail how to incrementally accelerate an OpenMP-parallelized CPU-only version of the RI-MP2 $\mathcal{O}(N^5)$-scaling kernel (\ref{mp2_energy} and \ref{eri_sym_decomp} combined) by means of OpenACC compiler directives~\cite{eriksen_openacc_rimp2_bookchapter_2016}. Herein, the (T) correction in \ref{ccsdpt_theory_subsection} is accelerated in a similar fashion, and the source code for both implementations (RI-MP2 and the (T) correction) accompany the present work under an MIT license~\cite{openacc_cc_codes}. Furthermore, we will here assess the potential of lowering the numerical precision in certain parts of the RI-MP2 and (T) kernels. In standard implementations of the two methods, all of the involved steps (evaluation of ERIs, amplitudes, fitting coefficients, etc., as well as the final energy assemblies) are performed using double precision arithmetics exclusively. However, motivated by two recent papers~\cite{vysotskiy_cederbaum_mixed_prec_rimp2_jctc_2011,knizia_werner_mixed_prec_rimp2_jctc_2011}, which claim that single precision arithmetics can be applied in the construction of the four-index ERIs (\ref{eri_sym_decomp}) used for RI-MP2 as well as in the construction of the triples amplitudes (\ref{analytical_triples_ampl}) for CCSD(T), we will do exactly this by substituting selective calls to {\texttt{dgemm}} with corresponding calls to {\texttt{sgemm}} in the codes. Not only is the performance bound to improve from the use of single precision arithmetics, but also the storage requirements on the GPU main memory will be lowered.

%
%
\section{Computational details}\label{com_details_section}

For the purpose of evaluating the implementations of the two methods in double or mixed floating point precision, we will conduct performance tests on chains ($\alpha$-helix arrangements) of alanine amino acids ([ala]-$n$, where $n$ is the number of residues) in cc-pV$X$Z ($X =$ D, T, Q, and 5) one-electron basis sets~\cite{dunning_1_orig} and corresponding cc-pV$X$Z-RI auxiliary basis sets~\cite{weigend_kohn_hattig_ri_basis_sets_jcp_2002,hattig_ri_basis_sets_5z_pccp_2005} (in the case of RI-MP2). Due to the difference in application range, the RI-MP2 and CCSD(T) implementations will be tested for different problem sizes (a composite measure of the number of electrons and number of basis functions); specifically, the RI-MP2 implementations are tested for the [ala]-6 system in the cc-pV$X$Z ($X =$ T, Q, and 5) basis sets as well as the systems [ala]-7 to [ala]-10 in a cc-pVQZ basis, whereas the CCSD(T) implementations are tested for the [ala]-1 system in the cc-pV$X$Z ($X =$ D, T, and Q) basis sets as well as the systems [ala]-2 to [ala]-6 in a cc-pVDZ basis.\\

In terms of computational hardware, the accelerators used are NVIDIA\textsuperscript{\textregistered} Kepler K40 GPUs (2880 processor cores $@$ 745 MHz (GPU Boost $@$ 875 MHz enabled for all calculations) and 12 GB main memory, from here on simply abbreviated as `K40s') and the host nodes are Intel\textsuperscript{\textregistered} Ivy Bridge E5-2690 v2, dual socket 10-core CPUs (20 cores $@$ 3.00 GHz and 128 GB main memory), i.e., {\texttt{OMP$\_$NUM$\_$THREADS}} $= N_{\text{threads}} = 20$ if not otherwise noted. The host math library is Intel\textsuperscript{\textregistered} MKL (version 11.2) and the corresponding device math library is CUBLAS (CUDA 7.5). All calculations are serial (non-MPI), and the OpenMP-/OpenACC-compatible Fortran compiler used is that of the PGI compiler suite (version 16.4). The Git hash of the code (Ref. \citenum{openacc_cc_codes}) used for the actual production runs is {\texttt{42f76337}}.\\

\begin{table}[h!]
\centering
\small
\caption{Comparison of correlation energies for the [ala]-6 and [ala]-7 systems (RI-MP2) and the [ala]-1 and [ala]-2 systems (CCSD(T)), using either no or $m$ number of GPUs (labelled CPU and GPU-$m$, respectively) in either double (DP) or mixed floating point precision (MP). The deviations (in $\mu\text{E}_{\text{H}}$) are reported with respect to corresponding reference results in double precision, and all mixed precision results are obtained as mean value results from five independent runs.}
\label{num_verification_table}
\begin{tabular}{l|r|rrrrrr}
  \hline
  \multicolumn{1}{c|}{Model} & \multicolumn{1}{c|}{CPU} & \multicolumn{1}{c}{GPU-1} & \multicolumn{1}{c}{GPU-2} & \multicolumn{1}{c}{GPU-3} & \multicolumn{1}{c}{GPU-4} & \multicolumn{1}{c}{GPU-5} & \multicolumn{1}{c}{GPU-6} \\
  \hline\hline
  \multicolumn{8}{c}{[ala]-6/cc-pVTZ/cc-pVTZ-RI} \\
  \hline
  RI-MP2 (DP) & $0.001$  & $0.001$ & $0.001$ & $0.001$ & $0.001$ & $0.001$ & $0.001$ \\
  RI-MP2 (MP) & $-0.049$ & $-0.614$ & $-0.664$ & $-0.706$ & $-0.744$ & $-0.740$ & $-0.758$ \\
  \hline
  \multicolumn{8}{c}{[ala]-7/cc-pVTZ/cc-pVTZ-RI} \\
  \hline
  RI-MP2 (DP) & $0.002$  & $0.002$ & $0.002$ & $0.002$ & $0.002$ & $0.002$ & $0.002$ \\
  RI-MP2 (MP) & $-0.060$  & $-0.854$ & $-0.902$ & $-0.956$ & $-0.956$ & $-0.970$ & $-0.992$ \\
  \hline
  \multicolumn{8}{c}{[ala]-1/cc-pVDZ} \\
  \hline
  CCSD(T) (DP) & $-0.002$  & $-0.002$ & $-0.002$ & $-0.002$ & $-0.002$ & $-0.002$ & $-0.002$ \\
  CCSD(T) (MP) & $-0.002$  & $-0.002$ & $-0.002$ & $-0.002$ & $-0.002$ & $-0.002$ & $-0.002$ \\
  \hline
  \multicolumn{8}{c}{[ala]-2/cc-pVDZ} \\
  \hline
  CCSD(T) (DP) & $0.004$  & $0.004$ & $0.004$ & $0.004$ & $0.004$ & $0.004$ & $0.004$ \\
  CCSD(T) (MP) & $0.005$  & $0.005$ & $0.005$ & $0.005$ & $0.005$ & $0.005$ & $0.005$ \\
  \hline
\end{tabular}
\end{table}

In all implementations, the tensors containing orbital energies, fitting coefficients, CCSD singles and doubles amplitudes, as well as ERIs are initialized with unique, yet arbitrary single and/or double precision numbers (depending on the context), as the calculation of these falls well outside the scope of the present study, cf. Ref. \citenum{eriksen_openacc_rimp2_bookchapter_2016}. This has been a deliberate choice on the part of the author, as {\bf{(i)}} it naturally modularizes the codes with a minimum of dependencies, making it easier for others to incorporate these into programs of their own should they wish to, and {\bf{(ii)}} the wall-time requirements will be determined solely from the computational steps outlined in \ref{rimp2_theory_subsection} and \ref{ccsdpt_theory_subsection}, an important (and necessary) requisite on the test cluster used in the preparation of the present work. However, the uniqueness of the input data makes for a safe monitoring and validation of the correctness of the CPU/GPU codes in-between related calculations; that the codes are indeed correct, also in a scientific sense, is reflected from the fact that, when provided with real HF and CCSD input data from an external program package (the {\textsc{lsdalton}} program~\cite{lsdalton,dalton_lsdalton_paper}), the energies agree accordingly, cf. \ref{num_verification_table}, to within the expected precision. Thus, albeit synthetic, the nature of the input data has no bearing on the results to follow in \ref{results_section}.

From \ref{num_verification_table}, we furthermore note how the (T) correction is significantly less sensitive to the use of mixed precision arithmetics (and the difference in architecture/math library between the CPU and the GPU) than the RI-MP2 method, a result which stems from the fact that only parts of the former are computed in single precision, as opposed to the latter, for which the entire algorithm, except for the final energy summation, is performed in single precision numbers, cf. \ref{implementations_section}. However, the results for both methods bear further testament to the validity of using single precision arithmetics in these types of calculations by supplementing the body of results in Refs. \citenum{vysotskiy_cederbaum_mixed_prec_rimp2_jctc_2011} and \citenum{knizia_werner_mixed_prec_rimp2_jctc_2011} by some for even larger system and problem sizes. Also, with respect to the new element here---the use of accelerators---we note how the (by all reasonable standards) marginal error with respect to the double precision reference increases slightly upon increasing the number of GPUs in the calculation; as we will see in \ref{results_section}, this result is closely related to the fact that the overall portion of the calculation, which is being handled by the GPUs, increases upon moving from one to six GPUs. We stress, however, that this hardware-related error remains perfectly tolerable as long as the inherent error from using single precision arithmetics is (i.e., on the order of $\leq10^{-6}$ a.u.).

%
%
\section{Results}\label{results_section}
\begin{figure}[htb]
        \centering
                \includegraphics[scale=0.80,bb=6 9 605 493]{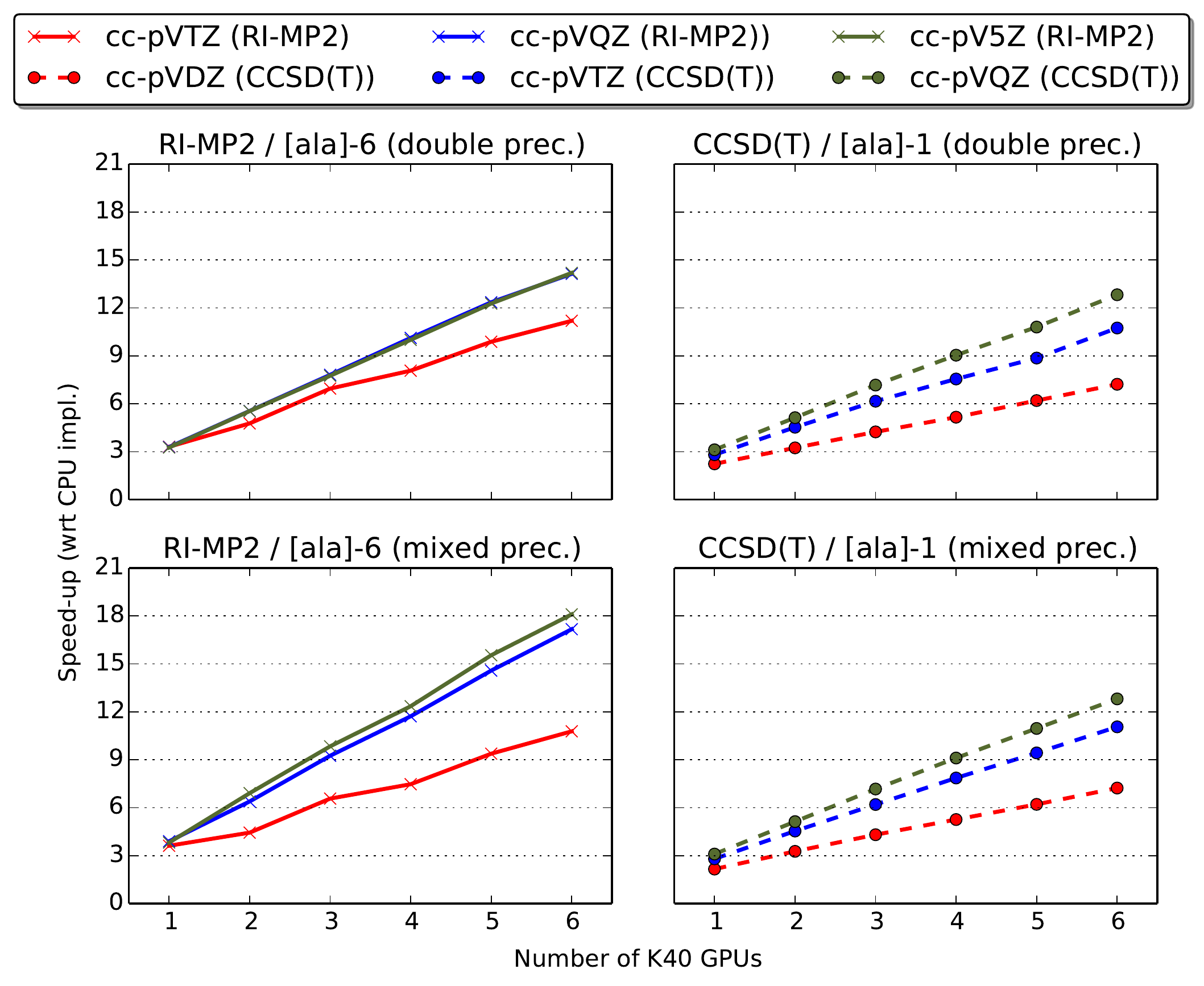}
        \caption{Multi-GPU scaling with the size of the basis set for the RI-MP2 and CCSD(T) implementations using double or mixed precision arithmetics.}
        \label{figure_1_fig}
\end{figure}
\begin{figure}[htb]
        \centering
                \includegraphics[scale=0.80,bb=6 9 573 493]{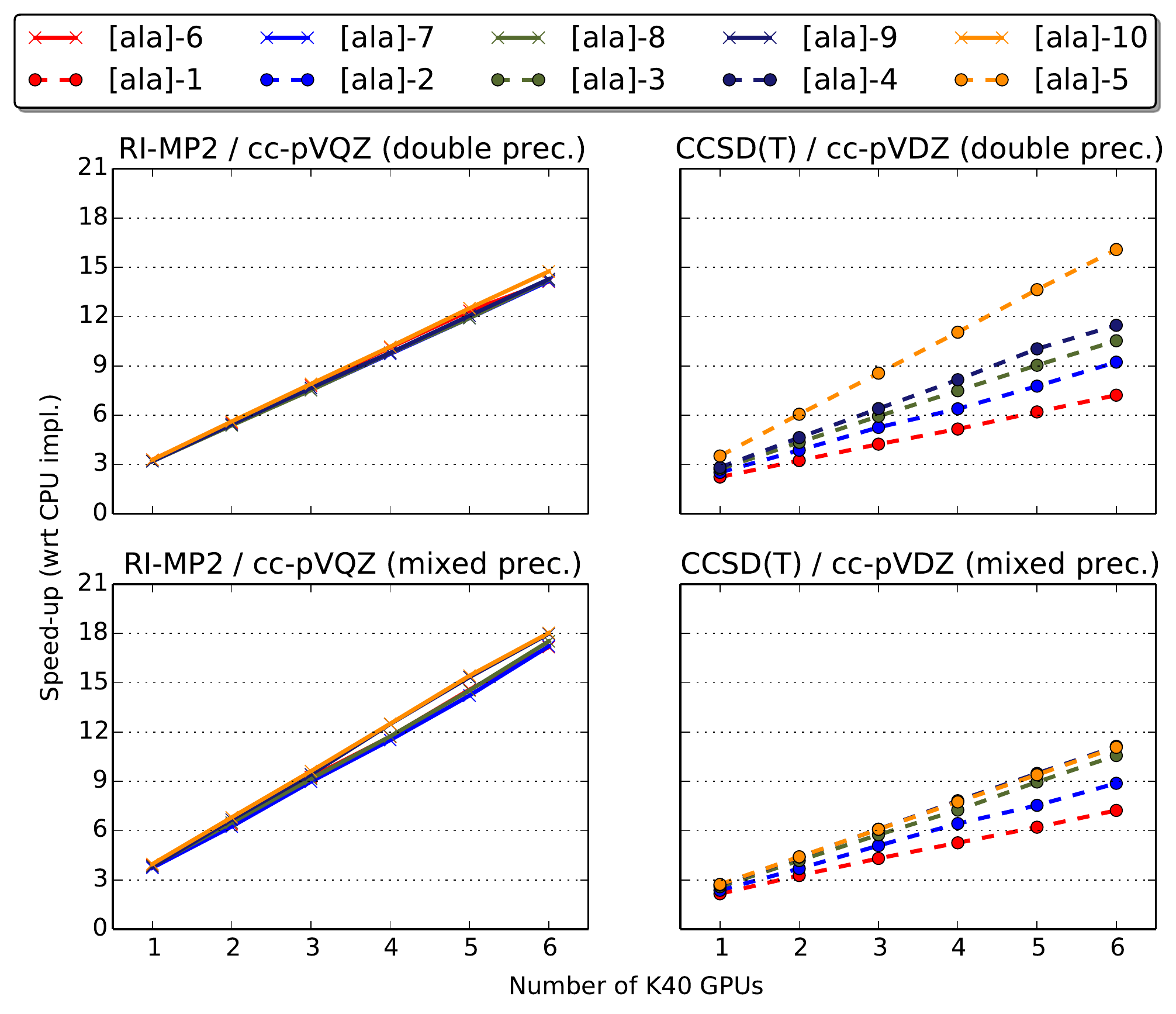}
        \caption{Multi-GPU scaling with the size of the molecular system for the RI-MP2 and CCSD(T) implementations using double or mixed precision arithmetics.}
        \label{figure_2_fig}
\end{figure}
In \ref{figure_1_fig} and \ref{figure_2_fig}, results for the total speed-up over the threaded CPU-only implementations (using a total of 20 OpenMP threads) are reported for fixed system size--varying basis sets (\ref{figure_1_fig}) and varying system sizes--fixed basis set (\ref{figure_2_fig}) as a function of the number of K40s involved in the calculations. Focusing first on the double precision RI-MP2 results, we note from \ref{figure_1_fig} how the [ala]-6/cc-pVTZ problem (by far the smallest of all the systems in the present RI-MP2 test) is too small for the use of the K40s to be optimal, while for all other combinations of system sizes and basis sets in \ref{figure_1_fig} and \ref{figure_2_fig}, the use indeed appears so ({\it{vide infra}}). From the double precision CCSD(T) results, on the other hand, a performance improvement is observed in both cases, with no obvious sign of saturation visible neither with respect to the size of the employed basis set in \ref{figure_1_fig} nor with respect to system size in \ref{figure_2_fig}. Moving from double to mixed precision arithmetics, the multi-GPU scaling of the RI-MP2 implementation is observed to be unaffected for the [ala]-6/cc-pVTZ problem, but significantly improved for all other tests; in particular, the performance is even improved in the transition from a quadruple- to a pentuple-$\zeta$ basis set in \ref{figure_1_fig}. For the mixed precision CCSD(T) implementation, however, the picture is much the same as for the equivalent implementation in double precision, although the improvement in scaling with increasing system size is now not as prominent. In explaining why this is so, we recall how only parts of the overall calculation of the (T) correction is performed using single precision arithmetics (the construction of the triples amplitudes), as opposed to the RI-MP2 implementation, in which only the final (computationally insignificant) energy assembly is performed in double precision numbers. Thus, the speed-up from calling {\texttt{sgemm}} over {\texttt{dgemm}} will be smaller. Also, the dimensions of the involved MMMs for the tested CCSD(T) problem sizes are greatly reduced with respect to those of the RI-MP2 problem sizes---a practical necessity, due to the larger memory and cost requirements of the former of the two methods---which further impedes (or disfavors) the scaling of the mixed precision implementation in the case of CCSD(T). The use of single precision does, however, allow for even larger problem sizes to be addressed, as we will see later on, since the overall CPU and GPU memory requirements are reduced. 
\begin{figure}[htb]
        \centering
                \includegraphics[scale=0.80,bb=7 9 577 470]{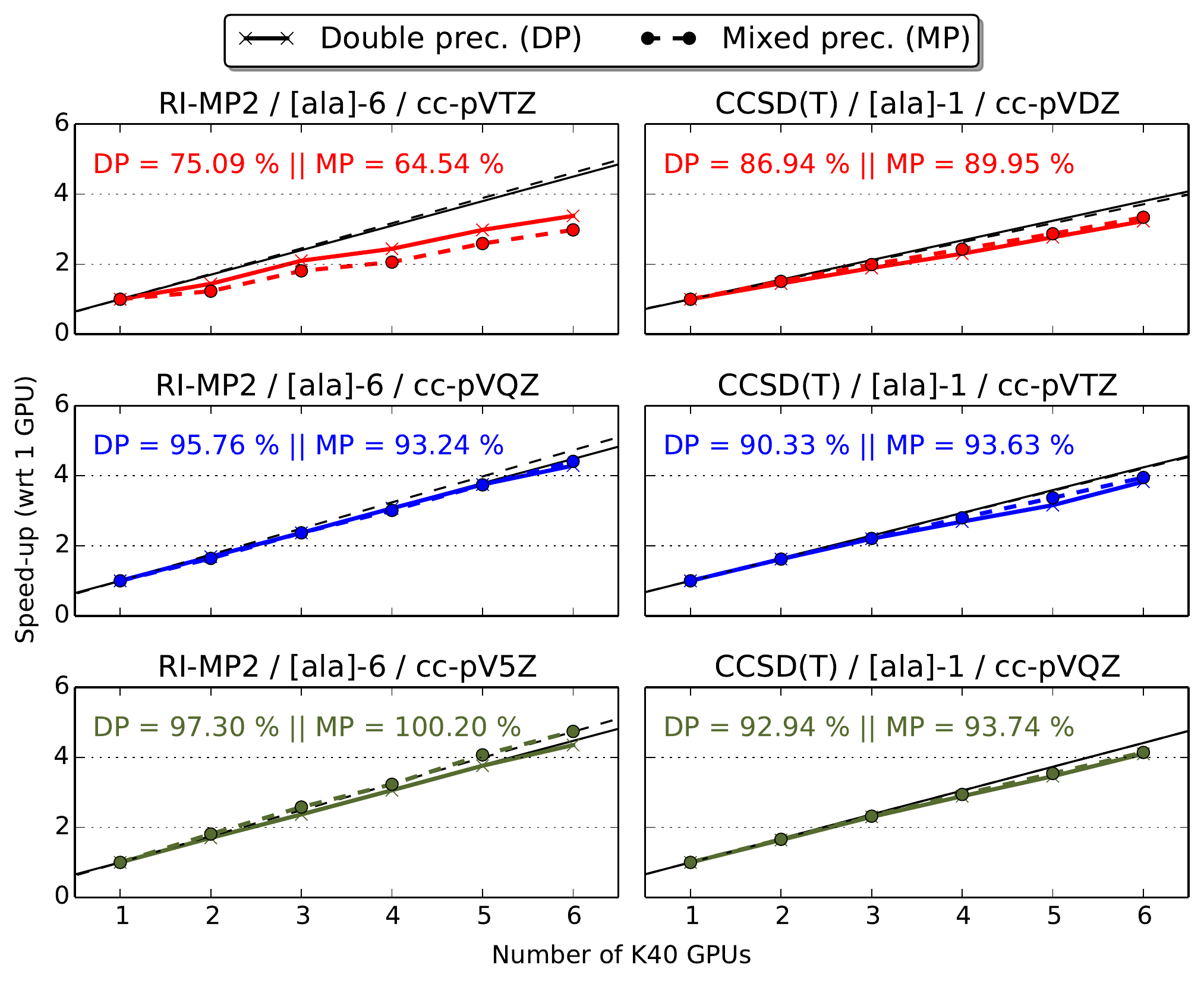}
        \caption{Deviation from ideal scaling with the number of K40s for combinations of a fixed system size (RI-MP2/[ala]-6 and CCSD(T)/[ala]-1) and varying basis sets (cc-pV$X$Z where $X =$ D, T, Q, and 5). For details, please see the text.}
        \label{figure_3_fig}
\end{figure}
\begin{figure}[htb]
        \centering
                \includegraphics[scale=0.80,bb=7 9 577 468]{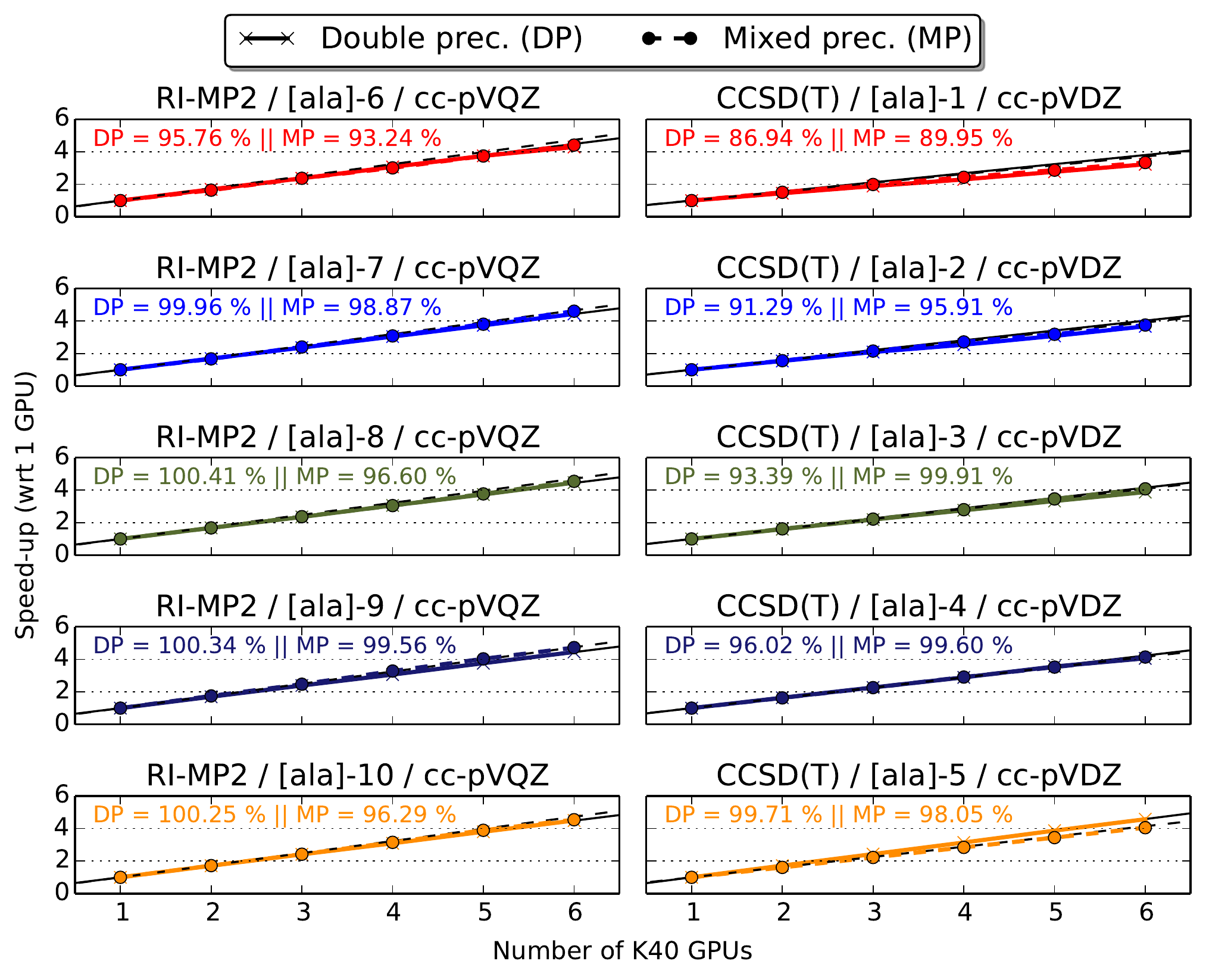}
        \caption{Deviation from ideal scaling with the number of K40s for combinations of varying system sizes ([ala]-$n$ where $n = 1-10$) and a fixed basis set (RI-MP2/cc-pVQZ and CCSD(T)/cc-pVDZ). For details, please see the text.}
        \label{figure_4_fig}
\end{figure}
Having assessed the accumulated speed-up over the ordinary CPU-only implementations, we next report results for the actual scaling with the number of K40s. In \ref{figure_3_fig} and \ref{figure_4_fig}, these scalings are presented, with the relative deviation from ideal behavior written out in the limit of six GPUs. As opposed to if the calculations were executed exclusively on the GPUs, i.e., in the non-hybrid limit where $N_{\text{threads}} == N_{\text{GPUs}}$, for which the ideal scaling is the trivial proportionality with the number of GPUs, $N_{\text{GPUs}}$ (performance doubling, tripling, etc., on two, three, etc., GPUs), this is not the case for hybrid CPU/GPU execution, i.e., whenever $N_{\text{threads}} > N_{\text{GPUs}}$, as each CPU core is now treated as an accelerator on its own. Thus, the ideal speed-up for the latter, heterogeneous case is defined as
\begin{align}
R = \frac{(N_{\text{threads}}-N_{\text{GPUs}})+N_{\text{GPUs}}S}{(N_{\text{threads}}-1)+S} \ . \label{r_speed_up}
\end{align}
In the definition of $R$ in \ref{r_speed_up}, the constant factor $S = N_{\text{threads}}K$, where $K$ is the time ratio between a CPU-only calculation ($N_{\text{threads}} =$ \texttt{OMP$\_$NUM$\_$THREADS}; $N_{\text{GPUs}} = 0$) and a GPU-only calculation using a single GPU ($N_{\text{threads}} = N_{\text{GPUs}} = 1$), accounts for the relative difference in processing power between a single CPU core (assuming ideal OpenMP parallelization on the host) and a single GPU. From the results in \ref{figure_3_fig} and \ref{figure_4_fig}, we note how the scaling becomes (near-)ideal for all but the smallest problem sizes, a statement valid for both models in both double and mixed precision, regardless of the fact that these dimensions are significantly reduced in the CCSD(T) tests, cf. the discussion above.

\begin{figure}[htb]
        \centering
                \includegraphics[scale=0.80,bb=6 8 603 469]{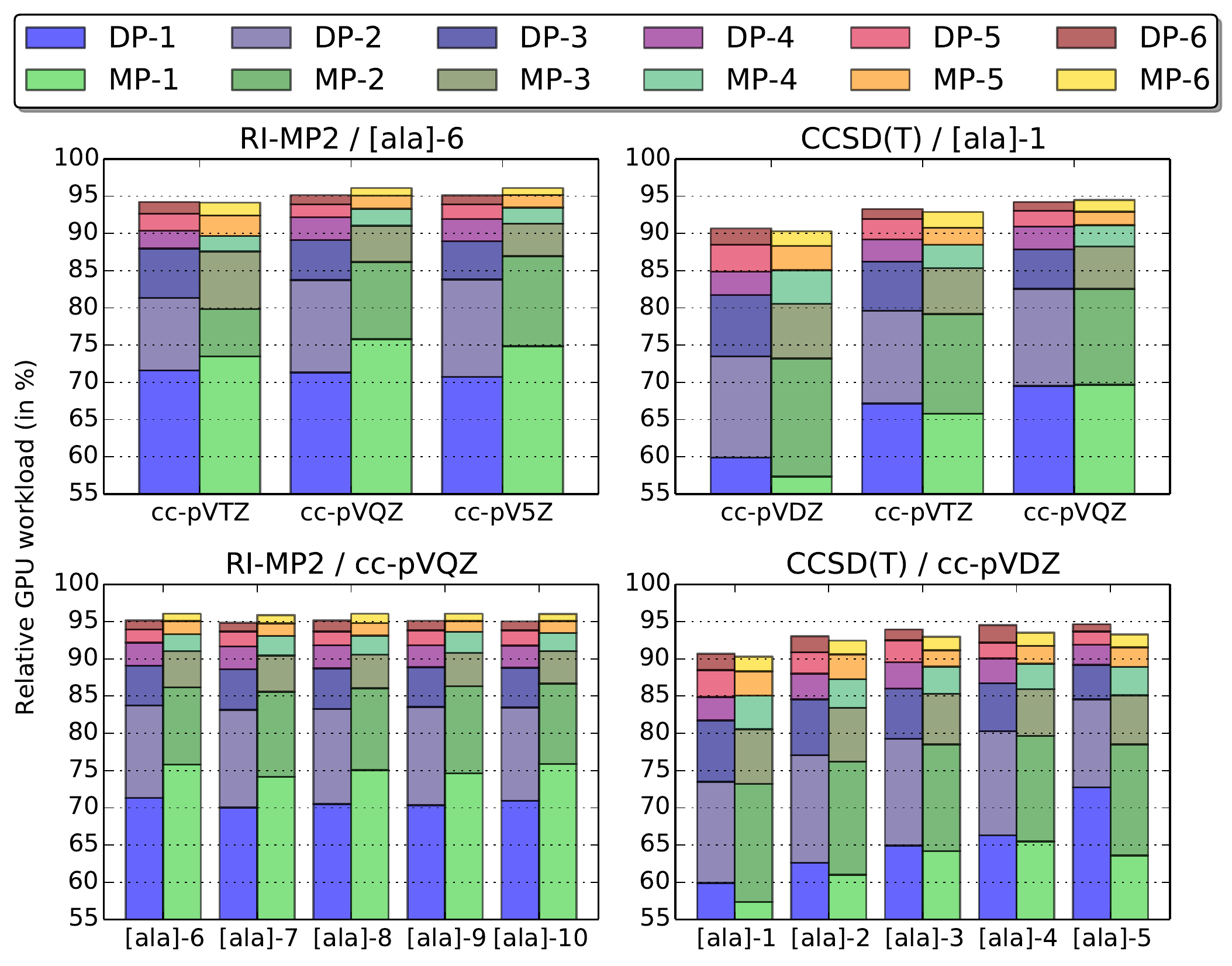}
        \caption{Accumulated relative GPU workload with number of K40s (DP-$n$/MP-$n$ where $n = 1-6$) using the implementations in either double (DP) or mixed precision (MP) numbers.}
        \label{figure_5_fig}
\end{figure}
In addition, we may monitor how large a percentage of the actual computations is being handled by the GPUs in the hybrid RI-MP2 and CCSD(T) implementations by noting how these involve a total of $T_{\text{RI-MP2}}$ and $T_{\text{CCSD(T)}}$ tasks of equal weight
\begin{subequations}
\begin{align}
T_{\text{RI-MP2}} &= N_{\text{o}}(1+N_{\text{o}})/2 \label{tasks_rimp2} \\
T_{\text{CCSD(T)}} &= \sum^{N_{\text{o}}-1}_{i=0}(N_{\text{o}}-i)(1+(N_{\text{o}}-i))/2 \label{tasks_ccsdpt}
\end{align}
\end{subequations}
where $N_{\text{o}}$ denotes the number of occupied orbitals. Through a dynamic OpenMP schedule, these individual tasks are distributed among the CPU cores and K40s in either of the implementations, so the scaling results in \ref{figure_3_fig} and \ref{figure_4_fig} may be complemented by corresponding results for the relative GPU workload. From the results, presented in \ref{figure_5_fig} as the accumulated workload in percent, we note how for the present problem sizes and CPU/GPU hardware, the actual utilization of the host node is minor (less than $10 \%$) when, say, three or more GPUs are attached to said node, regardless of the model. Still, the hybrid schemes are always faster than non-hybrid analogues, i.e., when $N_{\text{threads}} == N_{\text{GPUs}}$.

\begin{figure}[!ht]
        \centering
                \includegraphics[scale=0.80,bb=7 8 559 469]{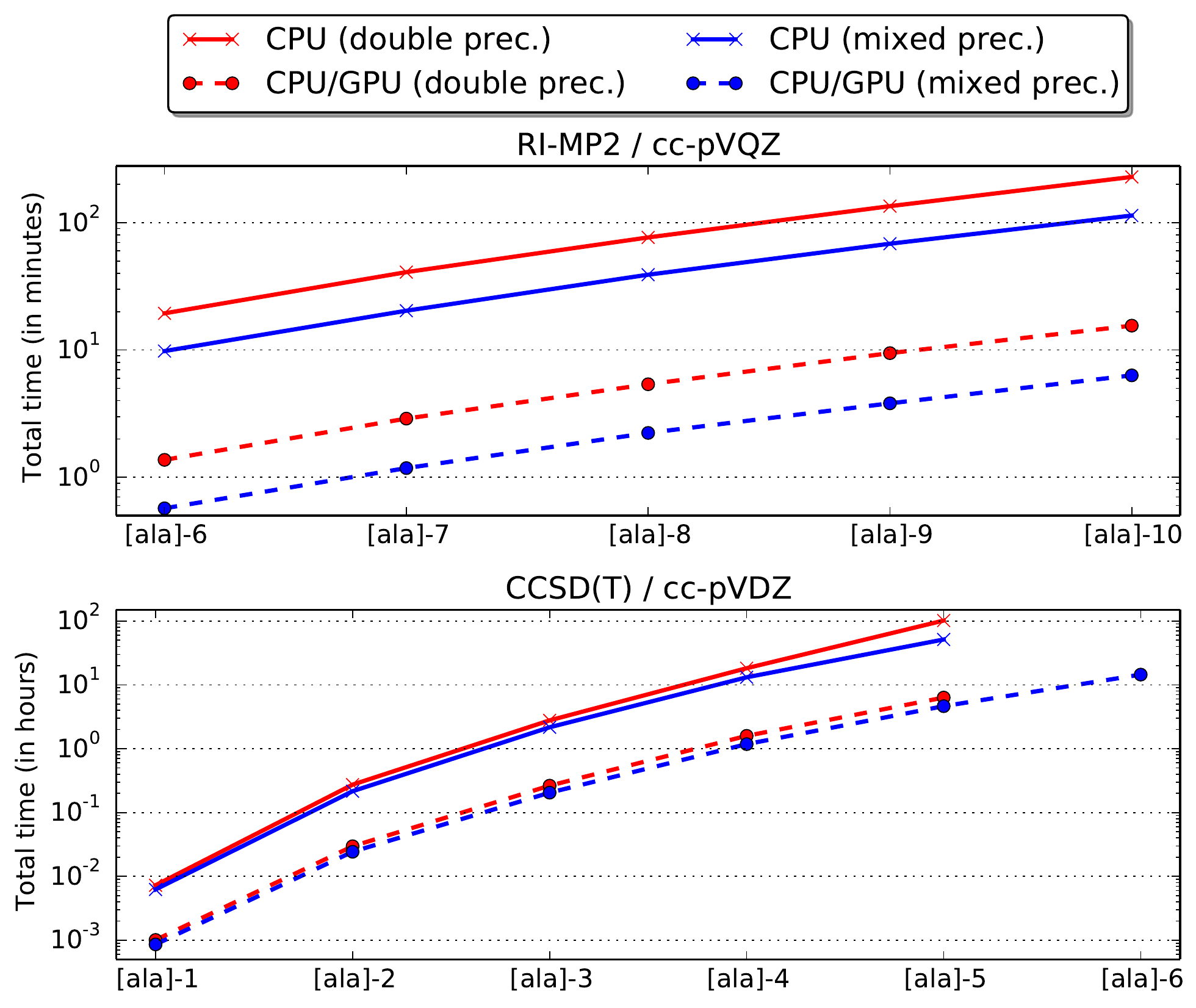}
        \caption{Total time-to-solution for the CPU-only and hybrid CPU/GPU implementations (using six K40s) of the RI-MP2 and CCSD(T) models.}
        \label{figure_6_fig}
\end{figure}
Finally, we compare the total time-to-solution for the CPU-only and hybrid CPU/GPU implementations of the RI-MP2 and CCSD(T) models in \ref{figure_6_fig}~\bibnote[ala_6_note]{Due to a limited amount of memory on the host node (128 GB), the accelerated CCSD(T) calculation on the [ala]-6/cc-pVDZ problem only spawned a total of 14 OpenMP threads instead of the default 20 threads. Out of these fourteen threads, six were devoted to the six K40s, while the remaining eight threads did computations on the host node. Finally, due to timelimit restrictions, the mixed precision CPU-only CCSD(T)/[ala]-6 calculation (using 14 OpenMP threads) was not performed, as were none of the corresponding [ala]-6 calculations in double precision since the memory requirements for these exceeded the size of the host node memory (i.e., the calculations required more than 128 GB).}. From these results, using six K40s, as in the present study, is seen to reduce the total time-to-solution over the CPU-only implementations---in either double or mixed precision---by at least an order of magnitude for all but the smallest possible problem sizes. This is indeed a noteworthy acceleration of both models. In particular, we note how for the RI-MP2 model, the accelerated calculation on the [ala]-10 system in a cc-pVQZ basis ($N_{\text{o}} = 195$, $N_{\text{v}} = 4170$, and $N_{\text{aux}} = 9592$ where $N_{\text{v}}$ and $N_{\text{aux}}$ are the number of virtual orbitals and auxiliary basis functions, respectively) took less time than the corresponding CPU-only calculation on the significantly smaller [ala]-6 system within the same basis ($N_{\text{o}} = 119$, $N_{\text{v}} = 2546$, and $N_{\text{aux}} = 5852$), while for the calculation of the (T) correction, an accelerated calculation on the [ala]-6/cc-pVDZ system ($N_{\text{o}} = 119$ and $N_{\text{v}} = 475$, cf. the note in Ref. \citenum{ala_6_note}) terminates in less time than a corresponding CPU-only calculation on the [ala]-4 system within the same basis ($N_{\text{o}} = 81$ and $N_{\text{v}} = 323$). On the basis of these results, it may be argued that the use of a combination of OpenMP and OpenACC compiler directives---as long as the complete fitting coefficients, integrals, CCSD amplitudes, etc., fit into main memory on the host---makes it somewhat unnecessary to explicitly parallelize the code using MPI with the complications that inevitably arise from this.

%
%
\section{Summary and conclusion}\label{conclusion_section}

In the present work, OpenACC compiler directives have been used to compactly and efficiently accelerate the $\mathcal{O}(N^5)$- and $\mathcal{O}(N^7)$-scaling rate-determining steps of the evaluation of the RI-MP2 energy and (T) triples correction, respectively. In the accelerated implementations, the kernels have all been offloaded to GPUs, and an asynchronous pipelining model has been introduced for the involved computations and data traffic. Due to their minimal memory footprints and efficient dependence on optimized math libraries, the implementations using either double or a combination of both single and double precision arithmetics are practically capable of scaling to as large systems as allowed for by the capacity of the host main memory.\\

In Ref. \citenum{song_martinez_gpu_ace_jctc_2016}, it was argued that one cannot completely rely on a general-purpose compiler in the search for better program transformations since such an imaginary object as an ideal compiler cannot exist, with reference to the somewhat demoralizing full employment theorem for compiler writers which states that for any optimizing compiler there will always exist a superior~\cite{appel_ginsburg_compiler_book_2004}. While this is perfectly true, instead of viewing it as a motivation for attempting to {\it{beat the compiler}} by creating additional compilation layers, such as code generators, testers, etc., one might take an alternative stand by obeying to a sort of ad hoc, yet generalizable philosophy of {\it{embracing the compiler}}; namely, given that an optimal coprocessor has not yet been mutually decided upon (for obvious reasons, as different vendors seek to promote their own variant), and given that consensus is yet to be reached on whether or not there is indeed a need for accelerators in the field of electronic structure theory, one might instead try to make the most out of what hardware is currently available (as well as near-future prospective hardware) by investing only the least possible amount of effort into porting one's code to this. The present work is intended to argue the case that the compiler directives of the OpenACC standard serve exactly this purpose by providing the means for targeting various coprocessors (e.g., GPUs or many-core x86 processors) in addition to the multi-core host node itself in an {\it{efficient}}, {\it{transparent}}, and {\it{portable}} high-level manner.\\

While the performance degradation of an OpenACC implementation (with respect to a hand-tuned low-level implementation) is bound to be larger for more complex electronic structure algorithms, such as the generation of ERIs or mean-field HF and DFT methods, than in the present case of MBPT methods, the application to the evaluation of the RI-MP2 energy and (T) triples correction is intended to illustrate a number of key advantages of the use of compiler directives over a reformulation of an optimized implementation for CPUs in terms of, e.g., CUDA or OpenCL for execution on GPUs. First and foremost, it is the opinion of the present author that accelerated code needs to be relatively easy and fast to implement, as new bottlenecks are bound to appear as soon as one part of a complex algorithm has been ported to accelerators (cf. Amdahl's law~\cite{amdahls_law_paper}). Second, the use of compiler directives guarantees---on par with the use of OpenMP worksharing directives for SIMD instructions on standard CPU architectures---that the final code remains functionally portable, i.e., the addition of accelerated code does not interfere with the standard compilation of the code on commodity hardware using standard non-accelerating compilers. Third, since the RI-MP2 and CCSD(T) methods alongside other, more advanced non-iterative CC many-body methods alike~\cite{kucharcki_ccsd_ptq_1_jcp_1998,*kucharcki_ccsd_ptq_2_jcp_1998,crawford_asymmetric_ccsdpt_ijqc_1998,piecuch_renormalized_cc_1,*piecuch_renormalized_cc_new_1,*piecuch_renormalized_cc_new_2,kallay_gauss_2005,*kallay_gauss_2008,ccsd_pert_theory_jcp_2014,*eom_cc_pert_theory_jcp_2014,*triples_pert_theory_jcp_2015,*quadruples_pert_theory_jcp_2015,*e_ccsd_tn_jcp_2016,*open_shell_triples_jcp_2016,*open_shell_quadruples_jcp_2016,*cc_conv_study_jcp_2016} are intrinsically reliant on large matrix-vector and matrix-matrix operations, the main need for accelerators in this context is for offloading exactly these. Thus, besides a number of small generic kernels, e.g., tensor index permutations or energy summations, compiler directives are primarily used for optimizing the data transfers between the host and the device(s), for instance by overlapping these with device computations. Hopefully, the generality of the discussion in the present work will encourage and possibly aid others to accelerate similar codes of their own. As a means to facilitate exactly this, the present implementations come distributed alongside this work for others to reuse in part or even in full~\cite{openacc_cc_codes}.\\

Finally, one particular potential area of application for the present implementations deserves a dedicated mentioning. While the discussion of the methods herein has been exclusively concerned with their standard canonical formulations for full molecular systems, we note how both methods have also been formulated within a number of so-called local correlation schemes, of which one branch relies on either a physical~\cite{kobayashi_dc_mp2_jcp_2007,mochizuki_fmo_mp2_cpl_2008,*ishikawa_fmo_rimp2_cpl_2009} or orbital-based~\cite{guo_cim_mp2_jpca_2014,*guo_cim_rimp2_scc_2014,friedrich_jcp_2007,*friedrich_incremental_mp2_ccsdpt_jctc_2009,marcin_dec_jcp_2010,*kasper_dec_jctc_2011,*ettenhuber_dec_orb_spaces_jcp_2016,*kjaergaard_dec_rimp2_cpc_2016} fragmentation of the molecular system. In these schemes, standard (pseudo-)canonical calculations are performed for each of the fragments before the energy for the full system is assembled at the end of the total calculation. Thus, by accelerating each of the individual fragment (and possible pair fragment) calculations, the total calculation will be accelerated as well without the need for investing additional efforts, and the resulting reduction in time-to-solution hence has the potential to help increase the range of application even further for these various schemes.

%
%
\section*{Miscellaneous}

The present work has also been deposited on the arXiv preprint server (arXiv:1609.08094).

%
%
\section*{Acknowledgments}

The author gratefully acknowledges support from the NVIDIA\textsuperscript{\textregistered} Corporation, in particular for granting access to an internal test cluster, the resources of which were used in the preparation of the present work. The author furthermore wishes to thank Mark Berger, Jeff Larkin, Roberto Gomperts, Michael Wolfe, and Brent Leback of the NVIDIA\textsuperscript{\textregistered} Corporation for general support and discussions, as well as Prof. T. Daniel Crawford of Virginia Tech, Dr. Filippo Lipparini of Johannes Gutenberg-Universit{\"a}t Mainz, and Dr. Radovan Bast of UiT The Arctic University of Norway for providing fruitful comments to the manuscript.

\newpage

\providecommand*\mcitethebibliography{\thebibliography}
\csname @ifundefined\endcsname{endmcitethebibliography}
  {\let\endmcitethebibliography\endthebibliography}{}


\begin{mcitethebibliography}{143}
\bibitem[ala()]{ala_6_note}
Due to a limited amount of memory on the host node (128 GB), the accelerated
  CCSD(T) calculation on the [ala]-6/cc-pVDZ problem only spawned a total of 14
  OpenMP threads instead of the default 20 threads. Out of these fourteen
  threads, six were devoted to the six K40s, while the remaining eight threads
  did computations on the host node. Finally, due to timelimit restrictions,
  the mixed precision CPU-only CCSD(T)/[ala]-6 calculation (using 14 OpenMP
  threads) was not performed, as were none of the corresponding [ala]-6
  calculations in double precision since the memory requirements for these
  exceeded the size of the host node memory (i.e., the calculations required
  more than 128 GB).\relax
\mciteBstWouldAddEndPunctfalse
\mciteSetBstMidEndSepPunct{\mcitedefaultmidpunct}
{}{\mcitedefaultseppunct}\relax
\EndOfBibitem
\end{mcitethebibliography}
\end{document}